\documentstyle[12pt]{article}
\input psfig

\newcommand{\eqref}[1]{Eq.~(\protect\ref{#1})}

\begin{document}

\begin{center}
{\bf {TWIST-STRETCH ELASTICITY OF DNA}}
\end{center}

\noindent C.S.~O'HERN, RANDALL D.~KAMIEN, T.C.~LUBENSKY, PHILIP NELSON\\
\noindent Department of Physics and
Astronomy, University of Pennsylvania,\\
\noindent Philadelphia, PA 19104\\

\bigskip

\noindent {ABSTRACT}

\bigskip

The symmetries of the DNA double helix require a new term in its
linear response to stress: the coupling between twist and stretch.
Recent experiments with torsionally-constrained single molecules give
the first direct measurement of this important material parameter. We
extract its value from a recent experiment of Strick, {\it et al.} and
find rough agreement with an independent experimental estimate
recently given by Marko. We also present a very simple microscopic
theory predicting a value comparable to the one observed.

\bigskip
\noindent {INTRODUCTION}
\bigskip

In this paper we will study the response of DNA to mechanical stress
using the methods of classical elasticity theory\cite{Kamien}.  While
many elements of DNA function require detailed understanding of
specific chemical bonds (for example the binding of small ligands),
still others are quite nonspecific. Moreover, since the helix repeat
distance of $l_0\approx3.4$nm involves dozens of atoms, it is
reasonable to hope that this length-scale regime would be long enough
so that the cooperative response of many atoms would justify the use
of a continuum, classical theory, yet short enough that the spatial
structure of DNA matters.

Since various important biological processes involve length
scales comparable to $l_0$ (notably the winding of DNA onto
histones), the details of this elasticity theory should prove
important. Yet, until recently, little was known about the relevant
elastic constants. Extensive experimental work yielded fair agreement
on the values of the bend and twist persistence lengths, though the
former was plagued with uncertainties due to the polyelectrolyte
character of DNA \cite{Record}. A simple model of DNA as a circular elastic
rod gives a reasonable account of many features of
its long-scale behavior, for example supercoiling
\cite{Benham}.

Recently, techniques of micromanipulation via optical tweezers and
magnetic beads have yielded improved values for the bend stiffness
from the phenomenon of thermally-induced entropic elasticity
\cite{smith}\cite{Bustamante}\cite{Marko1}, as well as
a direct measurement of a third elastic constant, the stretch
modulus\cite{smith2}\cite{Wang}. Significantly, the relation between
bending stiffness, stretch modulus, and the diameter of DNA turned out
to be roughly as predicted from the classical theory of beam
elasticity \cite{smith2}\cite{Wang}\cite{Landau}, supporting the
expectations mentioned above.

Still missing, however, has been any direct measurement of the elastic
constants reflecting the {\it chiral} ({\it i.e.} helical) character
of DNA. One such constant, a twist-bend coupling, was investigated by
Marko and Siggia \cite{Marko2}, but no direct experimental measurement
has yet been devised.  We will introduce a new chiral
coupling, the twist-stretch energy. Electrostatic effects do not
complicate the analysis of this coupling.  We will explain why our
term is needed, extract its value from the experiment of Strick, {\it
et al.} \cite{Strick}, and compare it to the prediction of a
microscopic model to see that its magnitude is in line with the
expectations of classical elasticity theory. J. Marko has
independently introduced the same coupling and estimated its value
from different experiments\cite{Markonew}; our values are in rough
agreement.

\newpage
\noindent {EXPERIMENT}
\bigskip

DNA differs from simpler polymers in that it can resist twisting, but
it is not easy to measure this effect directly due to the difficulty
of applying external torques to a single molecule.  The first
single-molecule stretching experiments constrained only the locations
of the two ends of the DNA strand.  The unique feature of the
experiment of Strick {\it et al.}  was its added ability to constrain
the {\it orientation} of each end of the molecule.

We will study Fig.~3a of ref.~\cite{Strick}. In this experiment, a constant
force of 8pN was applied to the molecule, and the end-to-end length
$z_{\rm tot}$ was monitored as the terminal end was rotated through $\Delta Lk$
turns from its relaxed state (which has $Lk_0$ turns). In this way the
helix could be over- or undertwisted by as much as $\pm10$\%. Over
this range of imposed linkage $z_{\rm tot}$ was found to be a linear function
of $\sigma$:
\begin{equation}
\epsilon= \epsilon_{\sigma = 0} -0.15\sigma,
\label{experiment}
\end{equation}
where $\sigma\equiv\Delta Lk/Lk_0$ and $\epsilon\equiv({\rm
z}_{\rm tot}/{\rm z}_{{\rm tot},0})-1$.  Thus $\sigma$ is the fractional excess
link, and $\epsilon$ is the extension relative to the relaxed state.
\eqref{experiment} is the experimentally
observed twist-stretch coupling.

\bigskip
\noindent {THEORY}
\medskip

\noindent \underline{Phenomenological Model}
\bigskip

A straight rod under tension and torque will stretch and twist. We can
describe it using the following reduced elastic free energy per
equilibrium length $z_{{\rm tot},0}$ of the rod:
\begin{equation}
f_{1}(\sigma,\epsilon)\equiv{F_1(\sigma,\epsilon)
\over k_BTz_{{\rm tot},0}}
={\omega_0^2\over2}\left[
\bar C \sigma^2+\bar B \epsilon^2
+2\bar D \epsilon\sigma\right] .
\label{f1}
\end{equation}
The twist persistence length is $\bar C \approx 75\,$nm \cite{Record},
while the helix parameter $\omega_0=2\pi/l_0=1.85$/nm.  We will take
$\bar B \approx1100 pN/\omega_0^2k_BT\approx 78$nm
\cite{Wang}. In the
experiment under study, there is an applied reduced force
$\tau=8$pN/$k_BT\approx 1.95/$nm. For a circular beam made of isotropic
material, the cross-term
$\bar D$ is absent \cite{Landau} because twisting is odd under spatial
inversion while stretching  is even. For a helical beam, however, we
must expect to find this term.

Setting $\tau = \partial f_1/\partial \epsilon |_{\sigma}$, we find
\begin{equation}
\epsilon=\epsilon_{\sigma=0} - {(\bar D/\bar B)}\sigma\ .
\label{compexperiment}
\end{equation}
Comparing to \eqref{experiment}, we obtain the desired result: $\bar
D=12\,$nm. To compare this to Marko's analysis, we note that his
dimensionless $g$ equals our $\bar D \omega_0$, so that we get
$g=22$. The rough agreement with Marko's result $g=35$ \cite{Markonew}
indicates that the data show a real material parameter of DNA and not
some artifact. We do not expect exact agreement, since Marko studied
the nonlinear overstretching transition of \cite{smith2}\cite{Cluzel};
our value came from the linear regime of small strains.

\bigskip
\noindent \underline{Microscopic Model}
\bigskip

\begin{figure}
\centerline{\psfig{figure=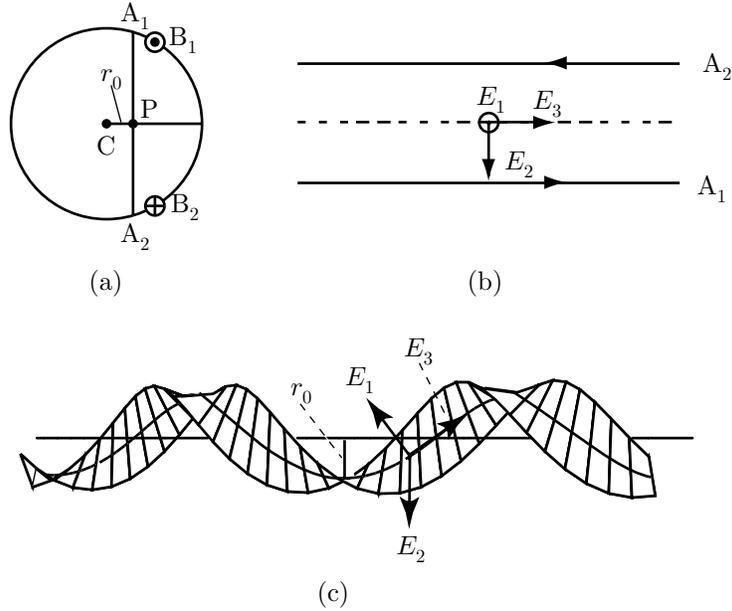}}
\caption{(a) Schematic representation of a cross section of DNA showing
its intersections (B$_1$ and B$_2$) with the phosphate backbones, its
intersection (line A$_1$PA$_2$) with the parameterizing helical
ribbon, and its helical center C. (b) A section of ribbon showing
oppositely oriented edges. At each point along the ribbon's center
(dotted curve), there is a triad of orthonormal vectors, ${\bf E}_1$,
${\bf E}_2$, and ${\bf E}_3$. ${\bf E}_3$ is parallel to the center
line, ${\bf E}_2$ points to one of the edges, and ${\bf E}_1$ is
perpendicular to the ribbon. (c) Representation of the helical
ribbon.}
\label{schematic}
\end{figure}

To gain further confidence in our result, we will now see how the
expected twist-stretch coupling emerges from a simple elastic model
for DNA. Fig.~\ref{schematic} reviews the relevant geometric properties of
DNA.  Base pairs are connected by inequivalent sugar-phosphate backbones whose
twisting pattern defines the major and minor grooves.  These
backbones intersect a given cross section of DNA at two points. Lines
parallel to the line connecting these two points sweep out ribbon-like
surfaces as the DNA twists.  We will parameterize the DNA structure by one of
these ribbons, whose center is a distance $r_0$ from the central helical
axis.  Any choice of $r_0$ is acceptable.  We will see, however, that a small
value of $r_0$ can explain the measured value of ${\overline D}$.

Our ribbon is described by the triad of unit vectors
${\bf E}_1$, ${\bf E}_3$, and ${\bf E}_2 = {\bf E}_3 \times {\bf E}_1$,
where ${\bf E}_3$ is the unit tangent vector to the center of the ribbon
and ${\bf E}_2$ points from the center of
the ribbon to one of its edges.
The triad varies as we move along the arc length $s$ of
the ribbon.  The motion is described by
\begin{equation}
{d {\bf E}_i \over ds} = -\epsilon_{ijk}\Omega_j {\bf E}_k .
\label{motion}
\end{equation}
The parameter $s$ labels each point along the central axis of the
ribbon in its unstressed state and runs from $0$ to $L$.  The actual
arc length along the distorted central axis of the ribbon will not be
$ds$ but rather $[1 + \epsilon_2(s)]ds$ where $\epsilon_2$ is the
intrinsic strain.  Therefore, the total length for constant $\epsilon_2$ is
$L^{'} = (1 + \epsilon_2) L$.  The intrinsic strain allows the
spacing between successive phosphate groups to change.

The edges of our ribbon, like the two sugar-phosphate backbones in DNA, are
distinguishable and point in opposite directions.  This symmetry can be
incorporated into our microscopic model by considering a rotation of
$180^{\circ}$ about the vector ${\bf E}_1$ followed by $s \rightarrow -s$.
Under this transformation ${\bf E}_2$ and ${\bf E}_3$ change sign, but $s$
derivatives of these vectors do not.  Also, ${\bf E}_1$ does not
change sign, but $ d {\bf E}_1/ds$ does.  Therefore, the free energy
should remain unchanged upon changing the sign of $\Omega_1$ but not
of $\Omega_2$ and $\Omega_3$\cite{Marko2}.  The most general reduced
free energy per length of ribbon
relative to that of the flat unstretched ribbon up to second order in
$\Omega_i$ and $\epsilon_2$ is
\begin{eqnarray}
f_{DNA} & = & {1 \over 2} [A^{'} {\Omega_1}^2 +
A(\Omega_2 - \Omega_{20})^2 + C (\Omega_3 - \Omega_{30})^2 +
B {\omega_0}^2 \epsilon_2^2 + \nonumber \\
& & 2D\omega_0 \Omega_3 \epsilon_2 +
2G\Omega_2\Omega_3 + 2K\omega_0 \Omega_2 \epsilon_2 - A {\Omega_{20}}^2
- C {\Omega_{30}}^2].
\label{dnaenergy}
\end{eqnarray}
$\epsilon_2$ does not change sign under $s \rightarrow -s$, and so it
can appear in combination with $\Omega_2$ and $\Omega_3$ in
\eqref{dnaenergy}.  This model is the simplest semi-microscopic model
that incorporates all of the symmetries of DNA.  It is an expansion to
harmonic order in first-order derivatives of the vectors ${\bf E}_i$
({\it e.g.}  of $\Omega_3 = -{\bf E}_1 \cdot {d {\bf E}_2 \over ds}$).
Thus, it is a model with quantitative predictive power so long as the
$\Omega_i$ are slow on a scale set by the distance $a = 0.6$~nm
between successive phosphate groups, ({\it i.e.} so long as $\Omega_i a \ll
1$).  In the ground state $\Omega_3 = \Omega_{30} \approx \omega_0 =
1.85/$nm, so that $\Omega_{30}~a \approx 1.1$ is not small.  This
implies that higher derivative terms ({\it e.g.} (${d\Omega_3
\over ds})^2$, etc.) are needed for a quantitative theory.  Nevertheless,
our simple semi-microscopic model captures the essential symmetry of the
DNA structure and allows us to address questions like the nature of the
twist-stretch coupling.

One can easily show that the center of the ribbon describes a helix
in the ground state of \eqref{dnaenergy}.
We will assume that $D$, $G$, and $K$ can be made small by an appropriate
choice of $r_0$.  Then, to keep the model as simple as possible, we will
simply set these parameters equal to zero for this choice of $r_0$.
We parameterize the helical ribbon using three angles
$\psi, \gamma,$ and $\phi$:
\begin{eqnarray}
{\bf E}_3 & = & \sin\gamma {\hat z} +
\cos\gamma {\hat \phi} \nonumber \\
{\bf E}_2 & = & \cos \psi ( - \sin \gamma {\hat \phi} + \cos \gamma
{\hat z} ) - \sin \psi {\hat \rho} \nonumber \\ {\bf E}_1 & = &-\sin \psi (
- \sin \gamma {\hat \phi} + \cos \gamma {\hat z} ) - \cos \psi {\hat
\rho} ,
\label{angles}
\end{eqnarray}
where ${\hat \rho}$ and ${\hat \phi}$ are cylindrical unit vectors
spinning at frequency ${\dot \phi}$.  If ${\dot \phi}$ is a constant
and $\psi = 0$, then the ribbon wraps around a cylinder of length $z_{\rm
tot}$ and radius $r$.
\begin{equation}
z_{\rm tot} = L(1 + \epsilon_2)\sin\gamma \hspace{0.2in} {\rm and}
\hspace{0.2in}
r = {\cos\gamma \over \omega}.
\label{length}
\end{equation}
In its ground state the helix has the following properties: $\Omega_1
= \Omega_{10} = 0$, $\Omega_2 = \Omega_{20} = \omega_0 \cos\gamma_0$,
$\Omega_3 = \Omega_{30} = \omega_0
\sin\gamma_0$, ${\dot \phi} = \omega_0 = \sqrt{{\Omega_{20}}^2 +
{\Omega_{30}}^2}$, and $\psi = \psi_0 = 0$.  Also, the ground state
length and radius of the molecule are, respectively, $z_{{\rm tot},0} =
L\sin\gamma_0$ and
$r_0 = \cos\gamma_0/ \omega_0$.

We now consider deviations in the ground state length $z_{{\rm tot}}$
and twist rate ${\dot \phi}$ of the molecule.  Since the total twist
is ${\dot
\phi} L$, the excess twist is $\sigma L$ where $\sigma =
({\dot \phi} - \omega_0)/\omega_0 = \delta {\dot \phi}/
\omega_0$.  Using \eqref{length} we
find that changes in length are produced both by intrinsic
strain $\epsilon_2$ and by changes in $\gamma$:
\begin{equation}
\epsilon = {z_{{\rm tot}} \over z_{{\rm tot}, 0}} - 1 =
\epsilon_1 + \epsilon_2
\hspace{0.2in} {\rm where} \hspace{0.2in} \epsilon_1 =
\cot\gamma_0 \delta \gamma .
\label{strain}
\end{equation}

The energy of harmonic deviations from equilibrium are obtained by
expanding $f_{DNA}$ to second order in $\delta \gamma(s)$, $\delta
\psi(s)$, and $\delta \phi (s)$.  The ground state is a periodic
helix implying these variables can be expressed in terms of Fourier modes
in different Brillouin zones defined by $\omega_0$.  Rotations of the
helix about space-fixed axes $x, y,$ and $z$ are described by
the variables $\delta \theta_x$, $\delta \theta_y$, and $\delta \theta_z$.
The variables $\delta \gamma$, $\delta \psi$, and $\delta \phi$
can be expressed in terms of these variables:
\newpage
\begin{eqnarray}
\delta \gamma & = & \cos\omega_0 s \, \delta \theta_x + \sin\omega_0 s \,\delta
\theta_y \nonumber \\
\delta \psi & = & {1 \over \cos \gamma_0}(\sin\omega_0 s \,\delta \theta_x -
\cos\omega_0 s \, \delta \theta_y ) \nonumber \\
\delta  \phi & = & \tan \gamma_0 ( \sin \omega_0 s \,\delta \theta_x - \cos
\omega_0 s \,\delta \theta_y ) + \delta \theta_z .
\label{spacefixed}
\end{eqnarray}
Thus, variations of $\delta \theta_x$ and $\delta \theta_y$ in the
first BZ give rise to variations in $\delta \gamma$ and $\delta \psi$
in the second BZ.  A complete long-wavelength theory can, therefore, be
expressed in terms of the first BZ components of $\delta \theta_x$,
$\delta \theta_y$, $\delta \gamma$, $\delta \psi$, and $\delta \phi$ (whose 1st
BZ component is equal to that of $\delta \theta_z$).

Using the relations for the $\Omega_i$ obtained from \eqref{motion} and
\eqref{angles}, we find
\begin{eqnarray}
\Omega_2-\Omega_{20} & = & \cos\gamma_0 \delta {\dot \phi} -
\omega_0\sin\gamma_0 \delta\gamma\ \nonumber \\
\Omega_3-\Omega_{30} & = & \sin\gamma_0 \delta {\dot \phi}
+ \omega_0\cos\gamma_0 \delta\gamma.
\end{eqnarray}
Then, using these expressions in $f_{DNA}$, integrating out $\psi$, and
remembering that length along the pitch axis is a factor of $\sin
\gamma_0$ less than the ribbon length, we find that the effective reduced
free energy per unit length of pitch axis is
\begin{equation}
f = f_B + f_{TS},
\label{longwavelength}
\end{equation}
where
\begin{equation}
f_{B} = {1 \over 4 \sin\gamma_0} ({A'} + A \sin^2\gamma_0 + C\cos^2\gamma_0)
({\dot\theta_x}^2 + {\dot \theta_y}^2) = {1 \over 2} {\overline
A}({\dot\theta_x}^2 + {\dot \theta_y}^2)
\label{bend}
\end{equation}
is the bending energy\cite{Marko3} and $f_{TS}$ is the twist-stretch energy
defined by
\begin{equation}
f_{TS} = {{\omega_0}^2 \over 2 \sin\gamma_0} [C_{\sigma \sigma} \sigma^2
+ B_{\epsilon_1 \epsilon_1} {\epsilon_1}^2 + B {\epsilon_2}^2 +
2 D_{\epsilon_1 \sigma} \epsilon_1 \sigma],
\label{coupling}
\end{equation}
where $C_{\sigma\sigma} = (A \cos^2\gamma_0 + C\sin^2\gamma_0)$,
$B_{\epsilon_1\epsilon_1} = (A {\sin^4\gamma_0 \over \cos^2\gamma_0} +
C \sin^2\gamma_0)$, and $D_{\epsilon_1\sigma} = (C -
A)\sin^2\gamma_0$.
The twist-stretch energy can be expressed in terms of the total
strain by setting $\epsilon_1 = \epsilon - \epsilon_2$ and
integrating over $\epsilon_2$.  The result is that $f_{TS}$
has the same form as \eqref{f1} with macroscopic elastic
constants  ${\bar B}$, ${\bar C}$, and ${\bar D}$
expressed in terms of our microscopic parameters
$A$, $B$, and $C$.
The stretch moduli associated with $\epsilon_1$ and $\epsilon_2$ add in
parallel to yield a total stretch modulus ${\bar B} = {1\over
\sin\gamma_0} ({1 \over B_{\epsilon_1 \epsilon_1}} + {1 \over B})^{-1}$.
The twist-stretch coupling is
\begin{equation}
{\bar D} = {B \over (B_{\epsilon_1 \epsilon_1} + B)} (C - A) \sin\gamma_0.
\label{dbar}
\end{equation}
These expressions for the twist modulus $C_{\sigma,\sigma}$, stretch
modulus ${\overline B}$, and twist-stretch modulus ${\overline D}$ are
valid for arbitrary values of the helix offset $r_0$.

Thus, a description of DNA in terms of a helical ribbon with an axis
offset from the central helical axis generates a twist-stretch
coupling even if the bare twist-stretch coupling ($D$ in
\eqref{dnaenergy}) is zero.  We can estimate the offset $r_0$
necessary to produce the measured $\overline D$ assuming it arises
entirely from \eqref{dbar}.  If we assume $A^{\prime} \approx A$ and
$\omega_0 r_0 \ll 1$, then $B \approx {\overline B} = 75$nm, $C
\approx {\overline C}= 78$nm, and $A \approx {\overline A} = 40$nm,
and we find ${\overline D}\approx (\omega_0r_0)^2 ({\overline
B}/{\overline A})({\overline C} - {\overline A})$ and $(\omega_0
r_0)^2 \approx 0.176$ or $r_0 \approx 0.23$nm\cite{Saenger}.
Corrections to this estimate are of order $(\omega_0 r_0 )^4 \sim
0.03$.  This result for the twist-stretch coupling ${\overline D}$ in
the limit of a small helix offset $r_0$ was found previously in
\cite{Kamien}.

\bigskip
\noindent {CONCLUSION}
\bigskip

We have modeled DNA as a thin helical ribbon and presented a complete
long-wavelength theory which includes energy costs due to bending,
stretching, and twisting the DNA molecule.  Using this theory, we were able
to relate semi-microscopic elastic constants to the experimentally
measured macroscopic elastic constants.  We have also calculated the
coefficient of the twist-stretch coupling and compared it to
torsionally constrained DNA stretching experiments.  We have found
that the experimental value of the twist-stretch coupling gives a
value $r_0$ for the radius of the helical ribbon in rough agreement
with the elastic center offset from the helix axis obtained from
crystallographic data\cite{Saenger}.

\bigskip
\noindent {ACKNOWLEDGMENTS}
\bigskip

\noindent We would like to thank
D. Bensimon, S. Block, and J. Marko for their help and for
communicating their
results to us prior to publication, and W. Olson for discussions.
RK, TL, and CO were  supported in part by NSF grant DMR96--32598.
PN was supported in part by NSF grant
DMR95--07366.

\bigskip

\end{document}